\documentclass[11pt,a4paper]{article}
\usepackage{jcappub}

\usepackage{chapterbib}    
\usepackage{color}         
\usepackage{graphics}      
\usepackage{lineno}

\title{\Large Search for 14.4 keV solar axions from M1 transition of $^{57}$Fe with CUORE crystals}

\author[1]{\large F.~Alessandria,}
\author[2]{R.~Ardito,}
\author[3,4]{D.~R.~Artusa,}
\author[3]{F.~T.~Avignone III,}
\author[5]{O.~Azzolini,}
\author[4]{M.~Balata,}
\author[4,6,7]{T.~I.~Banks,}
\author[8]{G.~Bari,}
\author[9]{J.~Beeman,}
\author[10,11]{F.~Bellini,}
% ...

\author{(The~CUORE~Collaboration)} 

\affiliation[1]{\small INFN - Sezione di Milano, Milano I-20133 - Italy}

\affiliation[2]{\small Dipartimento di Ingegneria Strutturale, Politecnico di Milano, Milano I-20133 - Italy}

\affiliation[3]{\small Department of Physics and Astronomy, University of South Carolina, Columbia, SC 29208 - USA}

\affiliation[4]{\small INFN - Laboratori Nazionali del Gran Sasso, Assergi (L'Aquila) I-67010 - Italy}

\affiliation[5]{\small INFN - Laboratori Nazionali di Legnaro, Legnaro (Padova) I-35020 - Italy}

\affiliation[6]{\small Department of Physics, University of California, Berkeley, CA 94720 - USA}

\affiliation[7]{\small Nuclear Science Division, Lawrence Berkeley National Laboratory, Berkeley, CA 94720 - USA}

\affiliation[8]{\small INFN - Sezione di Bologna, Bologna I-40127 - Italy}

\affiliation[9]{\small Materials Science Division, Lawrence Berkeley National Laboratory, Berkeley, CA 94720 - USA}

\affiliation[10]{\small Dipartimento di Fisica, Sapienza Universit\`a di Roma, Roma I-00185 - Italy }

\affiliation[11]{\small INFN - Sezione di Roma, Roma I-00185 - Italy }

% ...

\abstract{
We report the results of a search for axions from the 14.4 keV M1 transition from $^{57}$Fe in the core of the sun using the axio-electric effect in TeO$_{2}$ bolometers. The detectors are 5$\times$5$\times$5 cm$^{3}$ crystals operated at about 10 mK in a facility used to test bolometers for the CUORE experiment at the Laboratori Nazionali del Gran Sasso in Italy. An analysis of 43.65 kg$\cdot$d of data was made using a newly developed low energy trigger which was optimized to reduce the energy threshold of the detector. An upper limit of 0.58 c$\cdot$kg$^{-1}\cdot$d$^{-1}$ is established at 95\% C.L., which translates into lower bounds $f_{A}\geq3.12 \times 10^5$~GeV 95\% C.L. (DFSZ model) and $f_{A}\geq2.41 \times 10^4$~GeV 95\% C.L. (KSVZ model) on the Peccei-Quinn symmetry-breaking scale, for a value of $S=0.5$ of the flavor-singlet axial vector matrix element. These bounds can be expressed in terms of axion masses as $m_A\leq19.2$~eV and $m_A\leq250$~eV at 95\% C.L. in the DFSZ and KSVZ models respectively. Bounds are given also for the interval 0.35 $\leq S \leq$ 0.55.
}

\begin{document}
\maketitle


\linenumbers

\section{Introduction}
\label{sec:intro}

Quantum chromodynamics or QCD, largely accepted as the best theory describing strong interactions, contains one curious blemish known as ``the strong CP problem''.\ QCD predicts a large neutron electric dipole moment, of the order $|d_{n}|\approx10^{-16}$e$\cdot$cm, whereas the experimental bound is $|d_{n}|\leq2.9\times10^{-26}$e$\cdot$cm~\cite{bib:neutrondipole}.\
This fact puts an unnaturally small upper limit ($<10^{-10}$) to the $\theta_{QCD}$ parameter, the strength of the CP violating term present in the QCD.\ In order to explain this small value Roberto Peccei and Helen Quinn proposed~\cite{bib:pecceiandquinn} that the QCD Lagrangian possessed an additional global U(1) symmetry which modified the CP-violating term to:
\begin{equation}
\mathcal{L}_{\theta}=(\theta_{QCD}-\frac{a}{f_{A}})\frac{g_{s}^{2}}{32\pi^{2}}G^{\mu\nu}_{a}\tilde{G}_{a\mu\nu},
\label{eq:eq1}
\end{equation}
where $g_s$ is the strong coupling constant, $G^{\mu\nu}_{a}$ the gluon field, $\tilde{G}_{a\mu \nu}$ is its dual which violates CP symmetry, $a$ is a new pseudoscalar field, and $f_A$ is the Peccei-Quinn symmetry-breaking scale.\ 
Non-perturbative effects induce a potential for the field $a$ that has a minimum at $a=f_{A}\theta_{QCD}$ which causes the spontaneous breaking of the global U(1) symmetry.\ Later Weinberg~\cite{bib:weinberg} and Wilczek~\cite{bib:wilczek}, independently pointed out the properties of the Goldstone boson (the axion) resulting from the breaking of the U(1) symmetry. 
\par
The axion has a long history, with many theoretical and experimental papers published since the two seminal papers by Weinberg and Wilczek in 1978. Rather than attempt to review the subject, we refer the reader to comprehensive reviews by Raffelt \cite{bib:Raff}, Hagmann et. al \cite{bib:Hagg}, and Kim \cite{bib:KimRev}, and the many references therein.
We will just recall here that the ``standard'' Peccei-Quinn axion with a symmetry-breaking scale of the order of the electro-weak scale is ruled out by experiments. However, other models of  ``invisible'' axions which break the symmetry at much higher energies are still viable. The possibility that the axion might be most or part of the dark--matter has reinforced even further the interest for this field~\cite{bib:axion-dm}.

The purpose of the present work is to study the interaction of axions produced in the Sun with a terrestrial detector consisting of an array of TeO$_2$ bolometers operated underground and described in Section~\ref{sec:experiment}. The axion production mechanism is a competing branch of the M1 nuclear ground-state transition in $^{57}$Fe in the solar core. This first excited state at 14.4-keV is populated by thermal excitation as discussed in Section~\ref{sec:prod}. The detection mechanism relies on the axio-electric effect occuring in our TeO$_2$ detectors, which is the analogue of a photo-electric effect with the absorption of an axion instead of a photon. 

In this paper we will initially perform a totally model independent study, assuming that the coupling constants related both to the detection and production mechanisms are unconstrained free parameters. We will then focus on the non-hadronic axions that couple to gluons, electrons, and photons described in the model by Dine, Fischler, Srednicki, and Zhitnitski (DFSZ)\cite{bib:DFSZ}. In DFSZ or Grand-Unified Theory (GUT) model, axions couple to photons, gluons, and leptons at the tree level: this model is therefore quite appropriate for our experimental approach, which uses the coupling to electrons as a basic detection method. We will also consider another model by Kim and Shifman, Vainstein, and Zakharov \cite{bib:KSVZ}, the KSVZ or hadronic model, in which no coupling to leptons at the tree level occurs, but a weak radiatively induced coupling to electrons is possible due to axion's interaction with photons.

In both the DFSZ and KSVZ models, the mass of the axion, $m_A$, is directly related to the Peccei-Quinn symmetry-breaking scale, $f_A$, through the relation~(\ref{eq:eq2}) where $m_\pi=135$ MeV is the mass of the pion, $f_\pi \approx 92$ MeV the pion decay constant, while for the mass ratios $z=m_u/m_d=0.56$ and $w=m_u/m_s=0.029$,  $m_u$, $m_d$ and $m_s$ are the masses of the up, the down and the strange quark respectively\cite{CAST}:

\begin{equation}
m_{A}=\left(\frac{z}{(1+z+w)(1+z)}\right)^{\frac{1}{2}}\frac{f_{\pi}m_{\pi}}{f_{A}}=6~[\rm{eV}]{\Big(}\frac{10^{6}}{f_{A}~[\rm{GeV}]}{\Big)}.
\label{eq:eq2}
\end{equation}
 


\section{Axion-nucleon coupling: axion emission from $^{57}$Fe nuclei in the Sun}
\label{sec:prod}

As stated before, the axion source studied in this search is the M1 transition produced by thermal excitation of $^{57}$Fe in the solar core. The isotope $^{57}$Fe is stable and has 2.12$\%$ natural abundance, yielding an average $^{57}$Fe density in the Sun's core of $(9.0 \pm 1.2)\times 10^{19} \mathrm{cm}^{-3}$. The uncertainty is mostly due to the different metal diffusion models in the core and is computed in Ref.~\cite{serenelli}. The first excited state is at 14.4 keV, low enough to be thermally excited in the interior of the sun, which has an average temperature kT$\approx$1.3 keV \cite{Moriyama}\cite{haxtonlee}.  In this paper, we rely on Ref.~\cite{haxtonlee} for the determination of the expected axion flux. The 15\% error in the knowledge of $^{57}$Fe density is taken into account. The error in the $^{57}$Fe density was taken to be the error in the flux.

The Lagrangian that couples axions to nucleons is:
\begin{equation}
\mathcal{L}=a\bar{\psi}_{i}\gamma_{5}(g^0_{AN}\beta+g^3_{AN}\tau_{3})\psi.
\label{eq:eq4}
\end{equation}
Here, $g^0_{AN}$ and $g^3_{AN}$ are the iso-scalar and iso-vector coupling constants, model dependent, and $\tau_{3}$ is a Pauli matrix.

To compute the expected axion flux, the axion-to-photon branching ratio for the decay of the 1st excited state of  $^{57}$Fe  has to be taken into account \cite{frank_88, haxtonlee}:

\begin{equation}
\label{eq:branching1}
\frac{\Gamma_a}{\Gamma_{\gamma}}=\left(\frac{k_a}{k_{\gamma}}\right)^3\frac{1}{2\pi\alpha}\frac{1}{1+\delta^2}\left[\frac{g^0_{AN}\beta+g^3_{AN}}{(\mu_0-1/2)\beta+\mu_3-\eta}\right]^2
\end{equation}

where $\mu_0 =0.88$ and $\mu_3 =4.71$ are the isoscalar and isovector nuclear magnetic moments (in nuclear magnetons), $\delta$ is the E2/M1 mixing ratio for the  nuclear transition, $\beta$ and $\eta$ are nuclear structure dependent ratios. For the 14.4 keV de-excitation process their values are $\delta$ = 0.002, $\beta$ = -1.19 and $\eta$ = 0.8 \cite{haxtonlee}. With these values in Eq. (\ref{eq:branching1}), we have

\begin{equation}
\label{eq:branching2}
\frac{\Gamma_a}{\Gamma_{\gamma}}=\left(\frac{k_a}{k_{\gamma}}\right)^3 1.82 (-1.19g^0_{AN}+g^3_{AN})
\end{equation}

for an axion with a total energy of 14.4 keV. The resulting axion flux, given by \cite{bib:CAST_14keV}, will be: 
\begin{equation}
\label{eq:flux}
\Phi_a=\left(\frac{k_a}{k_{\gamma}}\right)^3 \times 4.56\times10^{23}(g^{{eff}}_{AN})^2 \quad \text{cm}^{-2}\text{s}^{-1} 
\end{equation}

in which the prefactor has not been approximated to 1 to account for the non-relativistic limit and where  $g_{AN}^{eff}\equiv(-1.19g_{AN}^{0}+g_{AN}^{3})$.

It is possible to evaluate the axion flux for specific models. In particular, in hadronic axions the axion-nucleon coupling constants are defined by the expressions \cite{frank_88,haxtonlee,bib:kaplan}:
\begin{equation} \label{eq:eq5}
\begin{split}
g^0_{AN} & =-7.8\times10^{-8} \left( \frac{6.2\times10^{6}\text{GeV}}{f_{A}} \right) \left( \frac{3F-D+2S}{3}\right) \\
g^3_{AN} & =-7.8\times10^{-8} \left( \frac{6.2\times10^{6}\text{GeV}}{f_{A}} \right) \left((D+F)\frac{1-z}{1+z}\right)
\end{split}
\end{equation}

where $F\approx0.48$ and $D=0.77$ are invariant matrix elements of the axial current\cite{frank_88}. The Spin-Muon Collaboration gives the range for the flavor-singlet axial vector matrix element $S$ of $0.15 \leq S \leq 0.50$ at 95\% C.L \cite{SpinMuon} while Altarelli \textit{et. al} give the range $0.37 \leq S \leq 0.53$ \cite{Altarelli}. For future considerations, we will consider the overlapping range $0.35-0.55$ for $S$.

It has to be underlined that values for $g_{AN}^{0}$ and for $g_{AN}^{3}$ in the DFSZ model depend on two additional unknown parameters, $X_u$ and $X_d$, the so-called Peccei-Quinn charges of the $u$ and $d$ quark respectively \cite{bib:kaplan}. These charges have positive-definite values constrained by the relation $X_u+X_d=1$. The fluxes in the DFSZ model lie within an interval whose lower and upper bounds are determined not only by the range assumed for $S$, but also on the different possible combinations of $X_d$ and $X_u$ values. The axion-nucleon coupling constants are defined in this case by the expressions~\cite{ bib:kaplan}:

\begin{equation} \label{eq:eq7}
\begin{split}
g^0_{AN} & =5.2\times10^{-8} \left( \frac{6.2\times10^{6}\text{GeV}}{f_{A}} \right)
\left( \frac{(3F-D)(X_u-X_d-3)}{6}+\frac{S(X_u+2X_d-3)}{3} \right)
\\
g^3_{AN} & =5.2\times10^{-8} \left( \frac{6.2\times10^{6}\text{GeV}}{f_{A}} \right)
\frac{D+F}{2} \left( X_u-X_d-3 \frac{1-z}{1+z} \right) \ .
\end{split}
\end{equation}

\begin{figure}[t]
\begin{center}
{\includegraphics[width=0.8\linewidth]{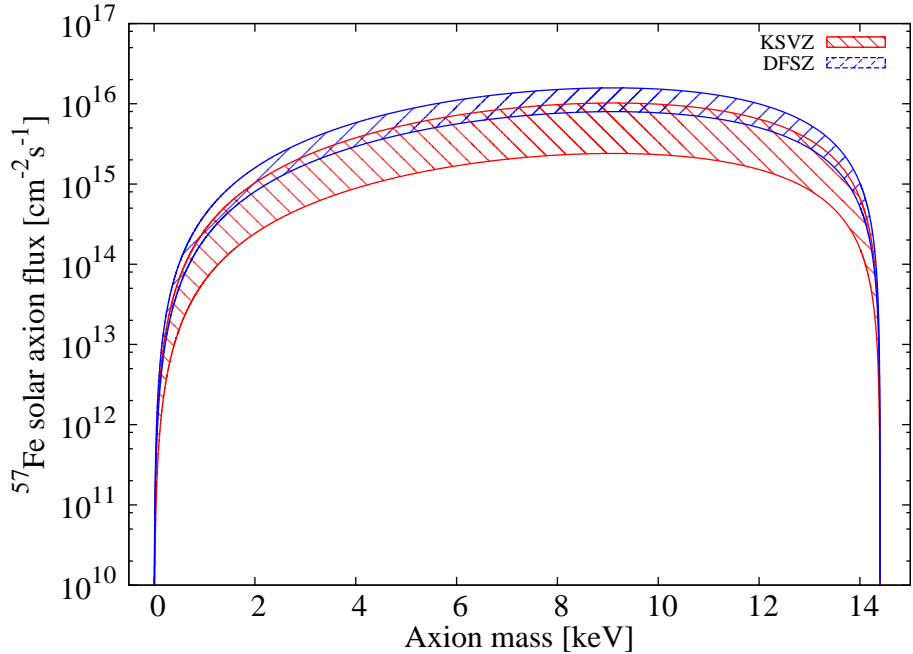}}
\end{center}
\caption{Expected fluxes of $^{57}$Fe solar axions as a function of the axion mass. The lower/red and the upper/blue regions (partially overlapping) refer to the KSVZ and DFSZ models respectively. The bands correspond to the ranges spanned by the parameters $S$,$X_d$ and $X_u$.}
\label{fig:fluxes}
\end{figure}

The fluxes of the 14.4 keV $^{57}$Fe solar axions have the same order of magnitude in the KSVZ and in the DFSZ model, as can be seen in Figure~\ref{fig:fluxes}. For a quantitative discussion of the data collected in the experiment here described, in the future we will assume $S=0.5$ and $X_d=1$ (and consequently $X_u=0$).

\section{Axion interaction with matter: the axio-electric effect}
\label{sec:axioelectric}
The detection mechanism used in the present work is the axio-electric effect, which is the equivalent of a photo-electric effect with the absorption of an axion instead of a photon: $A + e + Z  \rightarrow e+ Z$.

The purpose of this paper is to study $^{57}$Fe solar axions both in the relativistic and non-relativistic realms, with masses in the latter case approaching the 14.4 keV energy of the  $^{57}$Fe first excited state. It is very convenient therefore to choose an expression for the axio-electric cross section $\sigma_{Ae}$ which holds both in the extreme relativistic limit and for very cold axions, but is also capable of correctly describing the intermediate cases as well. Below we outline how this choice is made.

The two cross-section limits for $\beta \to 0$ and $\beta \to 1$, where $\beta$~is the axion velocity, have been computed by M.~Pospelov et al. (see for example Ref.~\cite{bib:Posp-2008}) in terms of $f_A$, and shown to be: 

\begin{equation} \label{eq:limits}
\begin{split} 
\sigma_{Ae} {\Big \vert} _{\beta \to 0} & \simeq  \sigma_{pe} (m_A) \frac {3 m_A ^2} {4 \pi \alpha f_A ^2 \beta} \\
\sigma_{Ae} {\Big \vert} _{\beta \to 1} & \simeq  \sigma_{pe} (E) \frac {E ^2} {2 \pi \alpha f_A ^2} \ , 
\end{split}
\end{equation}

where $\sigma_{pe}$ is the photoelectric cross section, $\alpha$ the fine structure constant, $E$ the axion energy and $m_A$ the axion mass.

We intend, however, to express the cross sections more generally in terms of a dimensionless coupling constant $g_{Ae}$, defined by the interaction Lagrangian

\begin{equation} \label{eq:lagr-gae}
\mathcal{L}_{int} = i g_{Ae} \overline{\psi} \gamma _5 \psi a
\end{equation}

which couples the axion field $a$ to the electron field $\psi$.

In order to do that, we recall that the formulae in Eq.~(\ref{eq:limits}) are obtained starting from the two equivalent forms~\cite{bib:Raff} of the axion-electron interaction Lagrangian

\begin{equation} \label{eq:lagr-fa}
\begin{split}
\mathcal{L}_{int1} & = - \frac {\partial _{\mu} a}{f_A} \overline{\psi} \gamma  ^{\mu} \gamma _5 \psi \\
\mathcal{L}_{int2} & = i \frac{2 m_e} {f_A} \overline{\psi} \gamma _5 \psi a \ .
\end{split}
\end{equation}

The comparison between Eq.~(\ref{eq:lagr-gae}) and the second expression in Eq.~(\ref{eq:lagr-fa}) shows that the two limiting forms for the axio-electric cross section in terms of $g_{Ae}$ can be achieved by replacing $f_A = 2 m_e / g_{Ae}$ in Eq.~(\ref{eq:limits}), obtaining: 

\begin{equation} \label{eq:limits-gae}
\begin{split}
\sigma_{Ae} {\Big \vert} _{\beta \to 0} & \simeq g_{Ae}^2 \sigma_{pe} (m_A) \frac {3 m_A ^2} {16 \pi \alpha \beta} \\
\sigma_{Ae} {\Big \vert} _{\beta \to 1} & \simeq g_{Ae}^2 \sigma_{pe} (E) \frac {E ^2} {8 \pi \alpha } \ . 
\end{split}
\end{equation}

A convenient general formula which reduces to the asymptotic expressions of Eq.~(\ref{eq:limits-gae}) has been proposed in Ref.~\cite{bib:derevianko, bib:derbin_gae} and adopted also in Ref.~\cite{bib:arisaka}. In the following analysis, we will use a similar expression, given by 

\begin{equation}
\label{eq:ae_cross_section}
\sigma_{Ae}(E)=\sigma_{pe} (E) \frac{g_{Ae}^2}{\beta}\frac{3E^2}{16\pi\alpha m_e^2}\left(1-\frac{\beta^{\frac{2}{3}}}{3}\right) \ ,
\end{equation}

which differs from the aforementioned formula -- reported in Ref.~\cite{bib:derevianko, bib:derbin_gae, bib:arisaka} -- as the exponent $\beta$ is $2/3$ instead of unity as in the quoted papers. We have introduced this change since, as suggested by M.~Pospelov~\cite{bib:posp3}, the modified formula in Eq.~(\ref{eq:ae_cross_section}) describes with better accuracy the cross section over the full $\beta$ range.

In specific axion models, the dimensionless constant $g_{Ae}$ is related to the electron mass and $f_{A}$ so that 

\begin{equation}
\label{eq:gae_gen}
g_{Ae} =  C_{e}m_{e}/f_{A} \ ,
\end{equation}

where $C_{e}$ is a model-dependent parameter that is of the order unity when the coupling to electrons occurs at the tree level. We have already seen that $C_e=2$ in the approach followed by M.~Pospelov et al.~\cite{bib:Posp-2008}.

In the DFSZ axion models, where the coupling at the tree level occurs, the parameter $C_e$ is usually expressed as

\begin{equation}
\label{eq:beta_dfsz}
C_e=\frac{1}{3} \cos ^2 (\beta _{DFSZ}) \ , 
\end{equation}

where tan$(\beta _{DFSZ})$ is the ratio of two Higgs vacuum expectation values. We will take $C_e=\frac{1}{3}$, which is equivalent to take $X_d=1$, where  $X_d$ is the Peccei-Quinn charge of the $d$ quark introduced before. Therefore, $g_{Ae}$ is numerically given by
\begin{equation} \label{eq:gaed}
g_{Ae} \simeq 1.68 \times 10^{-4} / f_{A}\text{[GeV]} \simeq 2.84 \times 10^{-8} m_A \text{[keV]}. 
\end{equation}

In the KSVZ axion model there is no tree-level couplings to electron, so $g_{Ae}$ is much smaller (by a factor of about $\alpha^{2}$), being determined only by radiative corrections \cite{bib:srednicki}:
\begin{equation}
g_{Ae}=\frac{3\alpha^{2}Nm_e}{2\pi
f_{A}}\left(\frac{E}{N}\ln\frac{f_{A}}{m_e}-\frac{2}{3}\frac{4+z+w}{1+z+w}\ln\frac{\Lambda}{m_e}\right),\label{Gaee}
\end{equation}
where $E/N=8/3$ in GUT models with $N=3$ (number of generations), and $\Lambda \approx 1$ GeV is the QCD cutoff scale. The factor containing $z$ and $w$ arises from the axion-pion mixing and is cut off at the QCD confinement scale. In this case, $g_{Ae}$ is numerically given by

\begin{equation} \label{eq:gaex}
g_{Ae} \simeq \frac {3.9 \times 10^{-8}}{f_{A}\text{[GeV]}} \left( 2.67 \times \ln \frac{f_A}{m_e} - 14.65 \right) \ . 
\end{equation}

\begin{figure}[t]
\begin{center}
{\includegraphics[width=0.8\linewidth]{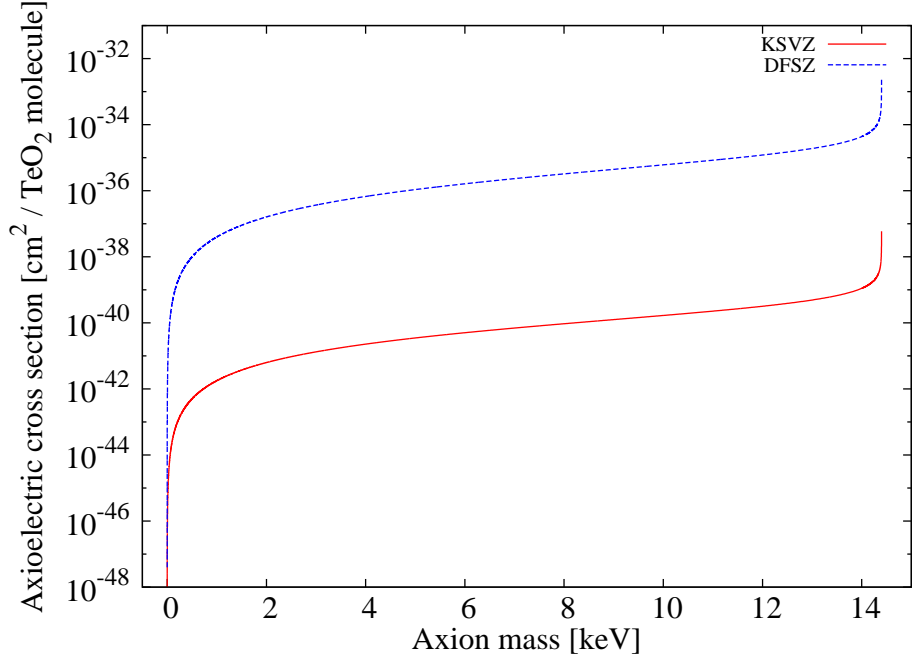}}
\end{center}
\caption{Cross sections on TeO$_2$ of 14.4~keV axions for axioelectric effect as a function of the axion mass in the case of KSVZ model (lower/red curve) and DFSZ model (upper/blue curve).}
\label{fig:cross}
\end{figure}

We have computed the cross sections foreseen by the two models in the case of a TeO$_2$ target and at the axion total energy of 14.4~keV. For the photelectric cross section on a TeO$_2$ molecule at 14.4~keV we have taken the value $1.17 \times 10^{-20}$~cm$^2$~\cite{bib:photoel}. The two cross sections are compared in Figure~\ref{fig:cross}, where the ratio of the order of $\alpha ^2$ between the KSVZ and DFSZ estimations is appreciable.

\section{Experiment}
\label{sec:experiment}

In the analysis shown in the present work we will focus on the results from a specific R\&D run dedicated to the CUORE experiment.

CUORE will be an array of 988 TeO$_2$ crystals, each with size $5\times5\times5$ cm$^{3}$ and weight 750 grams, which will be operated as bolometers at a temperature of about 10 mK to search for the neutrinoless double beta decay of $^{130}$Te and other rare events.\ A description of the CUORE technique and the basic principles behind bolometers is given in \cite{cuore1}, \cite{cuoricino}, \cite{cuoricino2}, and \cite{sensitivity}. 

The crystals to be used for CUORE are produced using special procedures developed to minimize radioactive contaminants~\cite{bib:crystal} at the Shanghai Institute for Ceramics of the Chinese Academy of Science (SICCAS), and are shipped in batches to the Laboratori Nazionali del Gran Sasso (LNGS) located in Assergi, Italy. Four crystals are taken at random from each batch to measure their radioactive contamination levels and evaluate their performance as bolometers at low temperature.\ Each one of these runs is called a Cuore Crystal Validation Run, or CCVR~\cite{bib:ccvr}.\ This paper will present the analysis data from the second run, known as CCVR2.

The four TeO$_{2}$ crystals (labeled B1,B2,B3,B4), have a total active mass of 3 kg and are mounted in a specially designed copper frame which is placed inside a dilution refrigerator at LNGS.\ The cryostat is maintained at a working temperature of approximately 8-10 mK throughout the duration of the run.

Glued on each CCVR bolometer are two Neutron Transmutation Doped (NTD) thermistors to read out the thermal signal~\cite{bib:ntd}. For CCVR2,  data were collected for a total  of 19.4 days for a total exposure of 43.65 kg$\cdot$d.\ Calibrations were performed in the middle and end of the run with a $^{232}$Th $\gamma$-ray source inserted inside the Pb shielding, close to the cryostat outer vacuum vessel.

The CUORICINO experiment \cite{cuore1,cuore2}, had bolometers whose typical threshold  was $\>$50 keV.\ Since the Q-value for $^{130}$Te is at 2527 keV \cite{qvalue, qvalue2,qvalue3}, a threshold at 50 keV was more than satisfactory for $\beta\beta$-decay searches.\ However, to investigate physical events at lower energies, new procedures were needed to lower the energy threshold.\ For this reason a special low--energy trigger was developed and applied offline, exploiting the fact that the standard CUORE DAQ used in the CCVR runs collects data continuously and with no hardware trigger using 125 Hz 18-bits digitizers. This trigger algorithm maximizes the signal to noise ratio by filtering the data with a known power spectrum and a reference signal shape. 
On each of the four bolometers, we selected the thermistor in which the trigger reached the lowest energy threshold.
The thresholds were 10 keV for B1, 3 keV for B2 and 2.5 keV for B3 and B4. For the present work we chose to include only the bolometers whose thresholds were lower than 4 keV as to include the peak at 4.7 keV. Although the physical interpretation of this line is unclear, the peak appeared to be constant in time \cite{bib:lowenergy} and therefore we used to monitor the stability of the cuts.
The efficiencies were  $0.91\pm0.10$ for B2 and $0.83\pm0.09$ for B3 and B4. We took the statistical fluctuation of the  4.7 keV peak rate in the energy spectrum as the systematic uncertainty of the efficiency. A full description of the trigger and its performance is given in \cite{bib:optimumtrigger,bib:lowenergy}.


Energy calibration in this very low energy region is not done with the $^{232}$Th $\gamma$-ray lines, which are normally used for higher energy calibration.\ The energy region between 2.5 and 300 keV is calibrated with a third order polynomial fit using a set of x-ray and $\gamma$-ray lines from metastable Te states that result from cosmogenic activation. The crystals spend several weeks above ground while they are being shipped by sea\footnote{The shipment of two (out of four) crystals was actually made by airplane, which induces a slightly higher activation.} between the production site in Shangai, China and arrival at the underground storage site at LNGS. The main $\gamma$-ray lines used in the calibration for present work are reported in Table \ref{tab:lines}.
\begin{table}[h]
\begin{center}
\begin{tabular}{|c|c|c|} \hline \hline
Energy [keV]   &    Source  &  Life-Time (days)  \\ \hline
30.4912                      &    Sb x-ray & -- \\
88.26 $\pm$ 0.08       &    $^{127m}$Te   &  109 $\pm$ 2  \\
105.50 $\pm$ 0.05       &    $^{129m}$Te   &  33.6 $\pm$ 0.1  \\
144.78 $\pm$ 0.03       &    $^{125m}$Te   &  57.40 $\pm$ 0.15  \\
247.5 $\pm$ 0.2       &    $^{123m}$Te   &  119.7 $\pm$ 0.1  \\
293.98 $\pm$ 0.04       &    $^{121m}$Te   &  154 $\pm$ 7  \\ \hline \hline
\end{tabular}
\end{center}
\caption{List of $\gamma$-ray lines from meta-stable Te isotopes used in this analysis for the energy calibration in the energy region between 2.5 and 300 keV. The lines are available from cosmogenic activation of Te during shipment, and have half-lives spanning from 33.6 days and 119.7 days.}
\label{tab:lines}
\end{table}
As a further check, the bolometers were irradiated with a $^{55}$Fe source deposited on the copper holder. The x-rays produced, with nominal energy between 5.888 and 6.490 keV, were shifted by only +(48$\pm$16) eV~\cite{bib:lowenergy}.

\section{Results}

In order to reject thermal and microphonic noise, the pulses are selected using the shape indicator variable described in \cite{bib:optimumtrigger}. This variable is based on the $\chi^2$ of the fit of the waveforms with the expected shape of the signal. Event selection is made by means of a scatter plot in which the two types of pulses form different bands. The determination of the  pulse shape cut efficiencies, described in detail elsewhere \cite{bib:lowenergy} have been evaluated on the 4.7 keV peak and are equal to 1. They were found to be almost completely energy independent for events above 3 keV.

\begin{figure}[t]
\begin{center}
{\includegraphics[width=\linewidth]{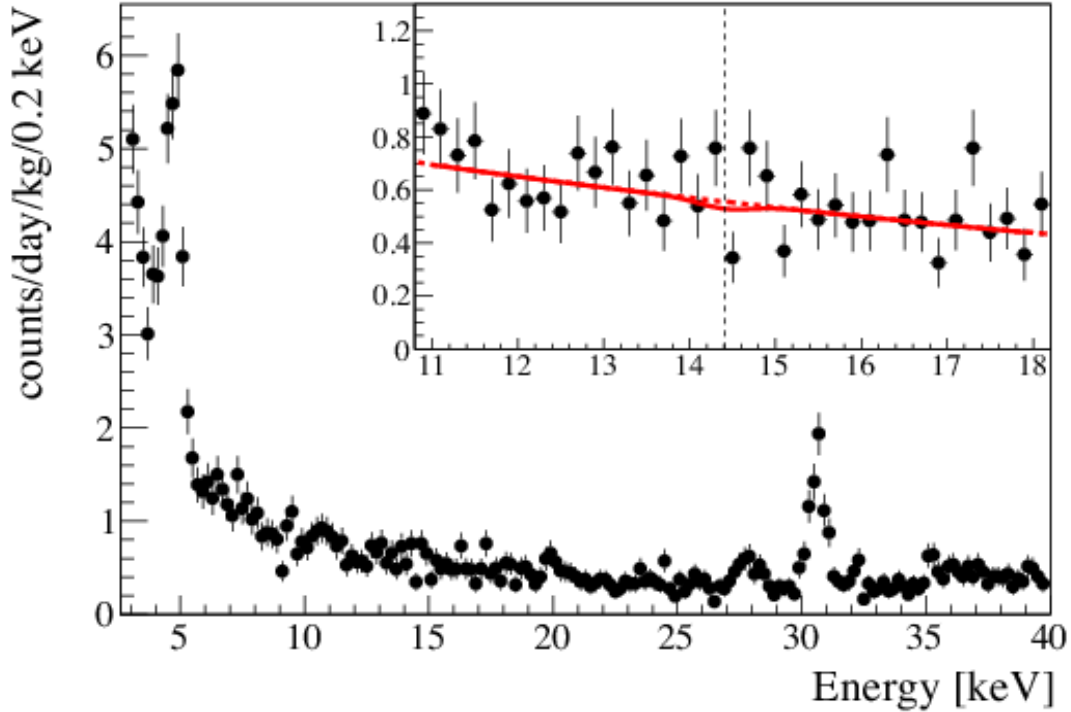}}
\end{center}
\caption{Energy spectrum for the low energy region between threshold and 40 keV. The two peaks at 4.7 keV and 30.49 keV are those used to study the energy resolution at low energy. The axion region around 14.4 keV is magnified in the inset with the fit result shown as red curve. No peak is observed at 14.4 keV, the M1 transition energy of $^{57}$Fe for solar axions. }
\label{fig:spectrum}
\end{figure}

The low energy spectrum, below 40 keV, is shown in Fig.\ \ref{fig:spectrum}.
It is modeled with two exponentials, one for the region lower than 5 keV, one
for the  background above 7 keV, and two Gaussians to model the peaks around 4.7 and 30.5 keV.
All parameters are free in the fit.
From the fit  to the spectrum in Figure \ \ref{fig:spectrum} we obtained consistent results for the two Gaussian widths:  $\sigma_{4.7}=0.29\pm0.02$ keV and  $\sigma_{30.5}=0.33\pm0.03$ keV. This was expected since at low energies the resolution of the bolometers is not dominated by signal fluctuations and we know that the energy dependence is weak. 
A fit in the region 11-18 keV (inset of Figure \ref{fig:spectrum}) is applied to extract limits on the axion detection rate.
We used an exponential function to model the continuum background and  a Gaussian centered at 14.4 keV for the axion peak.
The width of the Gaussian is fixed to the weighted average of the 4.7 and 30.5 keV resolutions $\sigma_{14.4}=0.31$ keV.
The best fit for the axion rate yields $-0.1\pm0.3$ c$\cdot$kg$^{-1}\cdot$d$^{-1}$. This is consistent with zero and the upper limit is set integrating the  posterior p.d.f  after marginalization over systematics uncertainties and with a flat prior on the rate in the physical region. Systematics consist of a 15\% uncertainties  on efficiencies and  $^{57}$Fe content in the solar core and are taken to be Gaussian. The resulting 95\% confidence level upper limit for the axion detection rate is 0.58 c$\cdot$kg$^{-1}\cdot$d$^{-1}$.

\section{Limits on the axion-relevant parameters}

Using the results described in the previous Section, it is possible to set excluded regions for the coupling involved in this search, i.e. $g_{Ae}$ and $g_{AN}^{eff} \equiv g^0_{AN}\beta+g^3_{AN}$. In the framework of specific models, we can also set limits on the axion mass $m_A$ or equivalently on the symmetry-breaking energy scale $f_A$.

Three scenarios will be considered:
\begin{enumerate}
  \item $g_{Ae}$ and $g_{AN}$ are completely model independent and the dependency on the axion mass comes only from kinematical factors (see Sections~\ref{eq:flux} and \ref{eq:ae_cross_section}).
  \item $g_{Ae}$ is a model independent parameter while $g_{AN}^{eff}$ is evaluated in the framework of the DFSZ and KSVZ models, using the considerations and the expressions reported in Section~\ref{sec:prod}.
  \item both $g_{Ae}$ and $g_{AN}^{eff}$ are evaluated in the framework of the DFSZ and KSVZ models, bringing in our knowledge of the axioelectric effect summarized in Section~\ref{sec:axioelectric}. 
\end{enumerate}

\begin{figure}[htbp]
\begin{center}
{\includegraphics[width=0.8\linewidth]{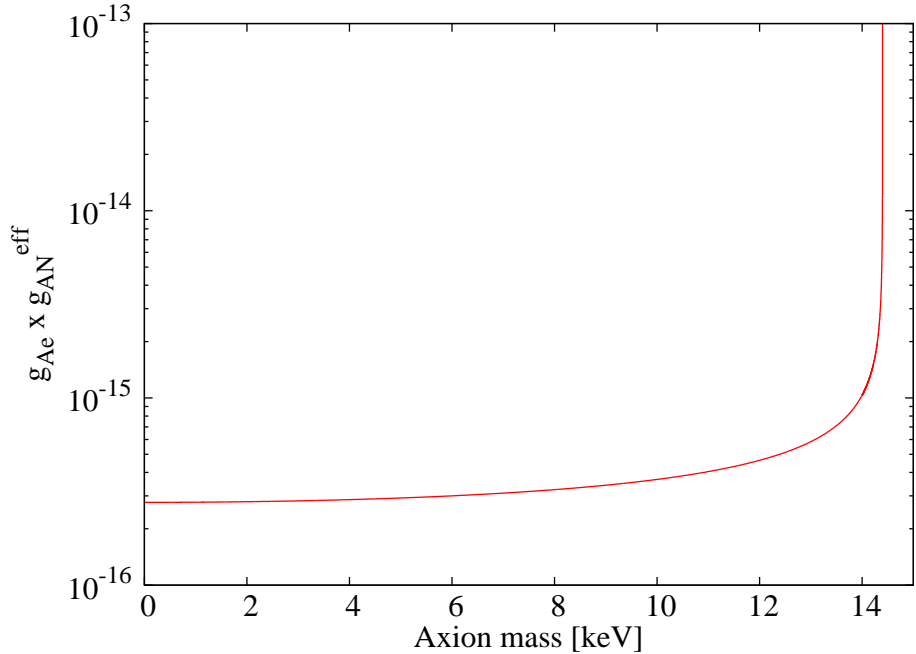}}
\end{center}
\caption{Upper limits for $ g_{Ae} \times g_{AN}^{eff} $ as a function of the axion mass $m_A$ obtained using the TeO$_2$ data.}
\label{fig:gae-gan}
\end{figure}

Using the relationships given in (\ref{eq:flux}) and (\ref{eq:ae_cross_section}), it is possible to obtain the upper limits for $(g_{AN}^{eff}) \times g_{Ae}$ as a function of the axion mass $m_A$. Fig. \ref{fig:gae-gan} shows this model independent limit.

\begin{figure}[htbp]
\begin{center}
{\includegraphics[width=0.8\linewidth]{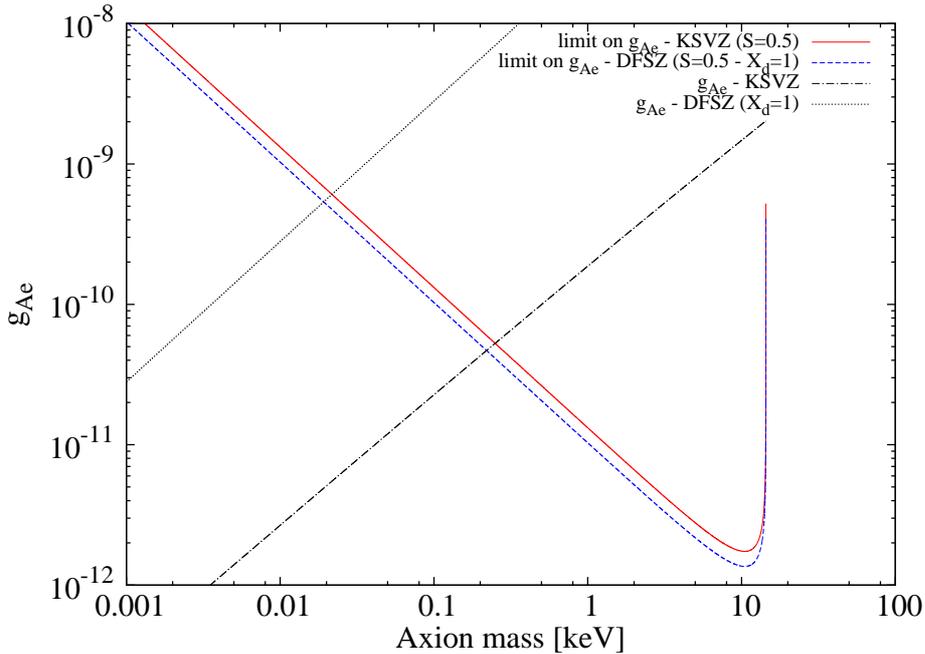}}
\end{center}
\caption{Upper limits on $g_{Ae}$ obtained using the TeO$_2$ results, assuming axion emission from the excited state of $^{57}$Fe in the in the KSVZ model (solid/red line) and in the DFSZ model (dashed/blue line), with $S=0.5$ and $X_d=1$ (see text). The inclined lines represent the relationships between the axioelectric cross sections and the axion mass in the KSVZ model (dashed-dotted line) and in the DFSZ model with $X_d=1$ (dotted line). The abscissae of the points where our bounds and the axiolectric predictions cross provide the limits on the axion mass (and consequently on the symmetry-breaking scale) in the two models (see text for these limits).}
\label{fig:gae}
\end{figure}

In the second approach, the allowed range of $g_{Ae}$ can be calculated as a function of the axion mass $m_A$, assuming the values predicted by the DFSZ and KSVZ models for the combination of $g_{AN}^0$ and $g_{AN}^3$ present in the branching ratio expression for the axion emission from the excited state of $^{57}$Fe. The curves reported in the plot assume $S=0.5$ for the flavor-singlet axial vector matrix element in both models, while the Peccei-Quinn charge of the $d$ quark $X_d$ is set equal to 1 for the DFSZ model. This is the value that maximizes the DFSZ axioelectric cross section that will be used to set a limit on the axion mass.  Figure~\ref{fig:gae} shows the restricted ranges for $g_{Ae}$.

Finally, we have considered the relationships between the axioelectric cross sections and the axion mass in the DFSZ model with $X_d=1$ and in the KSVZ model, and we have compared them with our bounds (see again Figure~\ref{fig:gae}). This allows to set limits on the axion masses, which correspond to 19.2 eV and 250 eV at 95\% c.l. in the DFSZ and KSVZ models respectively assuming $S=0.5$ in both cases. In terms of the symmetry-breaking energy scale $f_A$, these limits correspond to lower bounds of $3.12 \times 10^5$~GeV (DFSZ model) and $2.41 \times 10^4$~GeV (KSVZ model). The corresponding solar axion fluxes are $\sim 1.2 \times 10^{11}$ cm$^{-2}$s$^{-1}$ and $\sim 1.2 \times 10^{13}$ cm$^{-2}$s$^{-1}$. If we consider a range [0.35,0.55] for $S$ (as discussed in Section~\ref{sec:prod}), the limits on the masses are in the ranges [18.2,19.5]~eV in the DFSZ model and [232,339]~eV in the KSVZ model.  

\begin{figure}[htbp]
\begin{center}
{\includegraphics[width=0.6\linewidth]{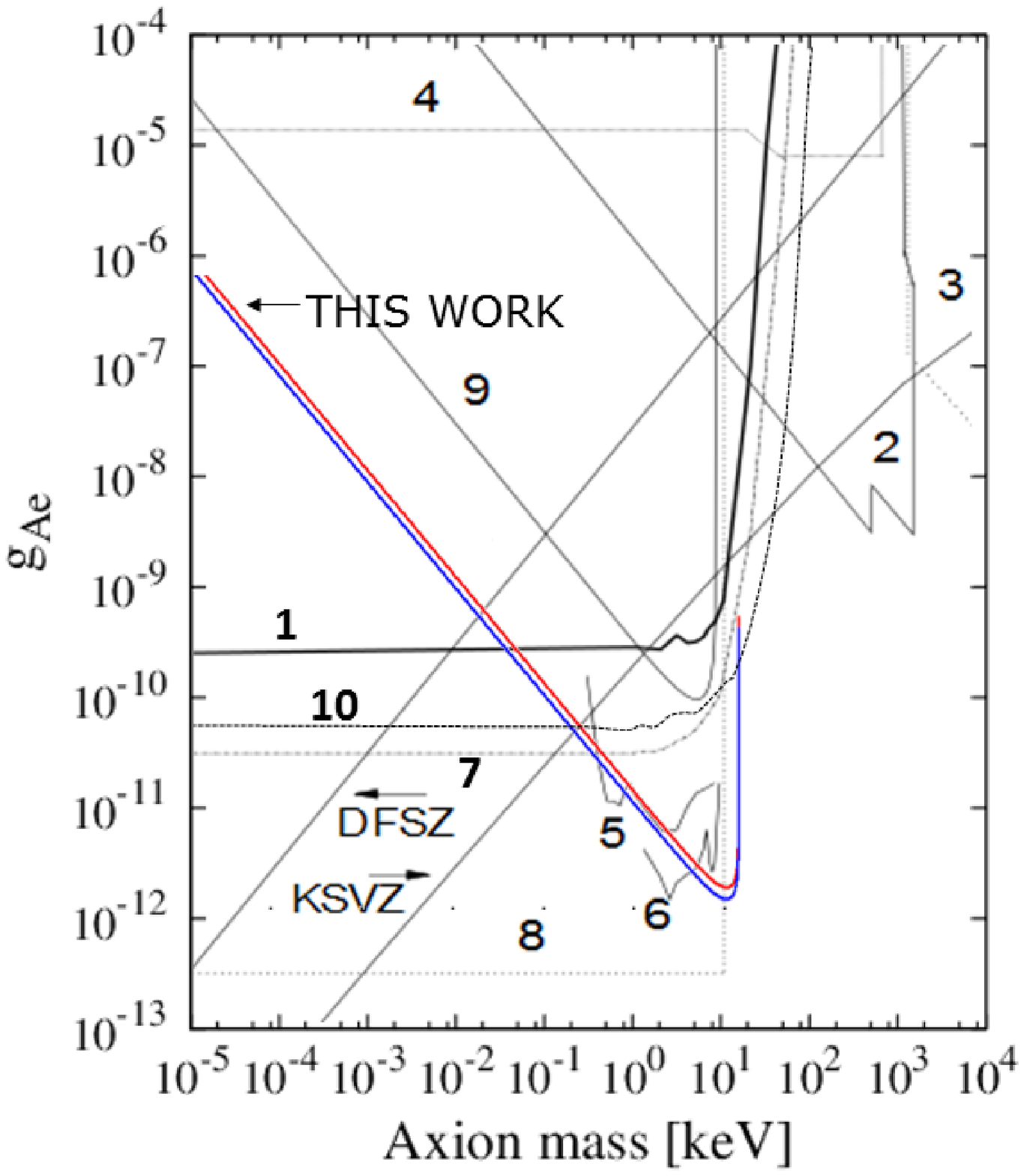}}
\end{center}
\caption{Bounds on $g_{Ae}$ obtained in this work (and already reported in Fig.~\ref{fig:gae}) are given by the upper (red) line and lower (blue) line for the KSVZ and DFSZ model respectively for the production of solar axions. These bounds are compared with (1) search for solar axions produced by Compton and bremsstrahlung processes with a Si(Li) detector~\cite{bib:derbin_gae}, (2) reactor experiments and solar axions with energy of 0.478 and 5.5 MeV~\cite{Altmann:1995, Chang:2007, bib:ctf-axions, bib:borex-axions, Derbin:2010}, (3) beam dump experiment~\cite{Konaka:1986, Bjorken:1988}, (4) decay of orthopositronium~\cite{Asai:1991}, (5) CoGeNT~\cite{Aalseth:2008}, (6) CDMS~\cite{Ahmed:2009}, (7) bound for the axion luminosity of the sun~\cite{Gondolo:2009}, (8) red giants~\cite{Raffelt:2008} , (9) experiment with $^{169}$Tm~\cite{Derbin:2011} and (10) XMASS~\cite{Abe:2012}. The inclined curves represent the relationships between $g_{Ae}$ and $m_A$ in the DFSZ and KSVZ models. The figure is adapted from Ref.~\cite{bib:derbin_gae}.}
\label{fig:comparison}
\end{figure}

Our results can be placed in a global context by comparing in a $g_{Ae}-m_A$~plane our bounds with those achieved by other searches, which use a variety of approaches and techniques. This comparison is shown in Fig.~\ref{fig:comparison}, from which one can appreciate that our results are at the level of the most competitive searches in this field.

\section{Summary and Conclusions}

An experimental search for axions emitted from the first excited state of  $^{57}$Fe in the solar core was performed. The calculation of the axion flux was made both in the framework of the DFSZ and KSVZ models. The detection technique employed a search for a peak in the energy spectrum at 14.4 keV when the axion is absorbed by an electron via the axio-electric effect. The cross section for this process is assumed to be proportional to the photo-electric absorption cross section of photons by electrons and is evaluated in the DFSZ and KSVZ axion models. In this experiment 43.65 kg$\cdot$d of data were analyzed resulting in a lower bound on the Peccei-Quinn energy scale of $3.12 \times 10^5$~GeV (DFSZ model) and $2.41 \times 10^4$~GeV (KSVZ model),  for a value for the flavor-singlet axial vector matrix element of $S=0.5$. 

The CUORE experiment will have about 740 kg TeO$_2$. With an anticipated live-time of 5 years the exposure will be $1.4\times10^{6}$ kg$\cdot$d. With a similar background as the one reported here the expected bound on $f_A$ will be increased by approximately an order of magnitude.    

\section{Acknowledgments}
The CUORE Collaboration thanks the Directors and Staff of the Laboratori Nazionali del Gran Sasso and the technical staffs of our Laboratories. This work was supported by the Istituto Nazionale di Fisica Nucleare (INFN); the Director, Office of Science, of the U.S. Department of Energy under Contract Nos. DE-AC02-05CH11231 and DE-AC52-07NA27344; the DOE Office of Nuclear Physics under Contract Nos. DE-FG02-08ER41551 and DE-FG03-00ER41138; the National Science Foundation under Grant Nos. NSF-PHY-0605119, NSF-PHY-0500337, NSF-PHY-0855314, and NSF-PHY-0902171; the Alfred P. Sloan Foundation; and the University of Wisconsin Foundation.
We also warmly thank C. Pe\~na-Garay, A. Serenelli and F. Villante for useful discussion about solar core $^{57}$Fe content.

\end{document}